\documentclass[aps,preprintnumbers,superscriptaddress,showpacs]{revtex4}
\usepackage{epsfig}
\usepackage{psfrag}
\usepackage{amsfonts}
\usepackage{amsmath}
\usepackage{graphicx}
\usepackage{dcolumn}
\usepackage{bm}

\begin{document}
\title{$R^4$ corrections to holographic Schwinger effect }

\author{Fei Li}
\affiliation{School of mathematics and
physics, China University of Geosciences(Wuhan), Wuhan 430074,
China}
\author{Zi-qiang Zhang}
\email{zhangzq@cug.edu.cn} \affiliation{School of mathematics and
physics, China University of Geosciences(Wuhan), Wuhan 430074,
China}

\author{Gang Chen}
\email{chengang1@cug.edu.cn} \affiliation{School of mathematics and
physics, China University of Geosciences(Wuhan), Wuhan 430074,
China}

\begin{abstract}
We consider $R^4$ corrections to the holographic Schwinger effect in
an AdS black hole background and a confining D3-brane background.
The potential between a test particle pair are performed for both backgrounds.
We find there is no potential barrier in the critical electric field,
which means that the system becomes catastrophically unstable.
It is shown that for both backgrounds increasing the inverse 't Hooft coupling parameter
$1/\lambda$ enhances the Schwinger effect. We also discuss the possible relation
between the Schwinger effect and the viscosity-entropy ratio $ \eta/s $ in strong coupling.
\end{abstract}


\maketitle
\section{Introduction}

The Schwinger effect is a non-perturbative phenomenon in quantum electrodynamics(QED),
which describes how the virtual electron-position pairs can become real particles
in a strong electric-field. The pair-production rate $\Gamma$ has been calculated
in the condition of the weak-coupling and weak-field by Schwinger \cite{Sch}.
Later, it was generalized to arbitary-coupling and weak-field by
Affleck-Alvarez-Manton (AAM) \cite{Aff}, via the relation
\begin{small}
\begin{equation}
\Gamma\sim exp\Bigg(-\frac{\pi m^2}{eE}+\frac{e^2}{4}\Bigg),
\end{equation}
\end{small}
where $\emph{m}$ is the electron mass, $\emph{e} $ is the elementary electric charge,
and $\emph{E}$ is the external electric field. Today, we know that this non-perturbative effect
is explained as a tunneling process and is not restricted to QED, but ubiquitously in
quantum field theories coupled to a U(1) gauge field. Thus, the Schwinger effect may
be an important tool to understand the vacuum structure and non-perturbative
aspects of string theory, and also quantum field theories.

Since Ads/CFT correspondence calls the duality  between the type IIB superstring theory
formulated on $AdS_5 \times S^5$ and $\mathcal{N}= 4$ SYM in four dimensions ( which realize
a construction that is coupled with the U(1) gauge field \cite{Mal,Gub,Aha} ), it is interesting
to consider the Schwinger effect in the context of the AdS/CFT correspondence.

One of the problems in the formula of AAM is that the critical value $E$ does not satisfy the
weak-field condition $eE \ll m^2$, where the $E$ is a certain value of the electric field
that makes the Schwinger effect occur without a tunneling process. In 2011, Semenoff and
Zarembo computed the production rate of the W bosons in the Coulomb branch
of $\mathcal{N} = 4$ SYM theory \cite{Sem}

\begin{small}
\begin{equation}
\Gamma\sim exp\Bigg[-\frac{\sqrt{\lambda}}{2}\Bigg(\sqrt{\frac{E_c}{E}}-
\sqrt{\frac{E}{E_c}}\Bigg)^2\Bigg],\qquad E_c = \frac{2\pi m^2}{\sqrt{\lambda}},
\end{equation}
\end{small}
where the value of critical field $E_c$ exactly coincides with the critical value of
the Dirac-Born-Infeld (DBI) action of the probe D3-brane with an electric field turning on.
$\lambda$ is the 't Hooft coupling constant. From the expression of production rate one
can anticipate the law of particle production to $\lambda$:

1) $E = E_c$, the production rate of particles $\Gamma \sim exp(0) = 1$ irrelevant to $\lambda$.

2) $E<E_c$ or $E>E_c$ , the production rate of particles $\Gamma \sim exp(-\frac{\sqrt{\lambda}}{2}\beta)$ decreases by increasing the $\lambda$ $(\beta = [\sqrt{E_c/E}-
\sqrt{E/E_c}]^2 > 0)$.

There are many attempts to address the Schwinger effect in this direction after Semenoff
and Zarambo's work. For instance, the potential in the holographic Schwinger effect has been
analyzed in \cite{Sat}. The holographic Schwinger effect in a confining D3-brane background
with chemical potential was studied in \cite{Zha}. One review about this topic was
shown in \cite{Kaw}. Ordinarily, due to the existence of the stringy effect, there are
many higher derivative corrections to the Schwinger effect. Although there is little knowledge
about the forms of higher derivative corrections, generic corrections can be expected
to exist when considering the vastness of the string landscape \cite{Dou}.

It is known that calculation of the holographic Schwinger effect is highly related to string theory,
so it is natural to consider various stringy corrections. The $R^2$ corrections to Schwinger effect
have been studied in \cite{Zha1}. In this research, we would like to study how $R^4$ corrections
affect the Schwinger effect. Besides, it was shown in \cite{Kov} that $\eta/s\geq 1/4\pi$ could be
violated in theories with $R^4$ corrections, which makes it very interesting to study the connection between
the shear viscosity and the Schwinger effect in these $R^4$ theories.

In this paper, we will study $R^4$ corrections to the Schwinger effect. The organization of
this paper is as follows. In section $\mathbf{2}$, the background with $R^4$ corrections is briefly
reviewed. In section $\mathbf{3}$, the potential analysis for the AdS black hole background with
$R^4$ corrections and the evaluation of the critical electric field from the DBI action is performed.
In section $\mathbf{4}$, the Schwinger effect in a confining D3-brane background with $R^4$ corrections
is studied as well. Section $\mathbf{5}$ is devoted to conclusion and discussion.

\section{Setup}

Let us focus on the contribution to the free energy F coming from the ${\alpha^\prime}^3R^4$,
string correction \cite{Gri,Fre,Par,Gro} and the supergravity action. In the Einstein frame,
using the convention including $(F_5)^2$ in the action and imposing self-duality after the
equations of motion are derived, the structure of the tree level type $\textrm{II}$B string
effective action will be as follows \cite{Gub1,Gro1}:
\begin{small}
\begin{equation}
\begin{split}
&S_{IIB}=\frac{1}{16\pi G_{10}}\\
&\cdot\int d^{10}x\sqrt{-g}
\Bigg(R-\frac{1}{2}(\partial\phi)^2
-\frac{1}{4\cdot5!}(F_5)^2+k e^{-\frac{3}{2}\phi}\mathcal{W}+...\Bigg) ,
\end{split}
\end{equation}
\end{small}
where $G_{10}$ is the ten-dimensional Newton constant, and $k = \frac{1}{8}\zeta(3)(\alpha^\prime)^3$.
The term $ \mathcal{W} $ depends only on the four copies of the Weyl tensor
\begin{small}
\begin{equation}
\mathcal{W}=C^{\alpha\beta\gamma\delta}C_{\mu\beta\gamma\nu}C_{\alpha}^{\rho\sigma\mu}C^{\nu}_{\rho\sigma\delta}+
\frac{1}{2}C^{\alpha\delta\beta\gamma}C_{\mu\nu\beta\gamma}C_{\alpha}^{\rho\sigma\mu}C^{\nu}_{\rho\sigma\delta}.
\end{equation}
\end{small}

For the purpose of computing the corrected Sch-winger effect one can use the Kaluza-Klein reduced
five-dimensional action \cite{Buc}
\begin{small}
\begin{equation}
S=\frac{1}{16\pi G_{5}}\int d^{5}x\sqrt{-g}\Bigg(R+\frac{12}{R^2}+k \mathcal{W}\Bigg) ,
\end{equation}
\end{small}
where R is the radius of curvature of $ AdS_5 $, and $\mathcal{W}$ is given by Eq. (4) in five-dimensions. $k$
is the expansion parameter, which is connected with the 't Hooft coupling constant $\lambda$ in
$ \mathcal{N} = 4 \textrm{SYM} $ by
\begin{small}
\begin{equation}
k=\frac{\zeta(3)}{8}\lambda^{-\frac{3}{2}}\sim0.15\lambda^{-\frac{3}{2}}.
\end{equation}
\end{small}

Because the higher derivative terms are treated as perturbations in the equations of motion,
the reliable results will be restricted to small(or large) values of parameter $k$(or $\lambda$). Therefore, the DBI analysis and the SUGRA background are not accurate enough when the value of $\lambda$ extends to its smaller interval.
 The values for the 't Hooft coupling $\lambda$ were assumed as follows \cite{Fad}
\begin{small}
\begin{equation}
\lambda=\{8,12,20,100\},
\end{equation}
\end{small}
Since the correction of the inverse 't Hooft coupling parameter 1/$\lambda$ corresponds to the $\alpha^\prime$ correction in string theory \cite{Paw}, the dependence of the $\alpha^\prime$ correction effect on parameter $\lambda$ may be studied. Theoretically, the $\alpha^\prime$ correction effect decreases gradually as the 't Hooft coupling parameter $\lambda$ increases, and it approaches zero as $\lambda$ approaches infinity.

The function of the $\alpha^\prime$-corrected metric is \cite{Gub1,Fad}
\begin{small}
\begin{equation}
ds^2
=-\frac{r^2(1-w^{-4})}{R^2}T(w)dt^2
+\frac{r^2}{R^2}X(w)(d{x^i})^2+
\frac{R^2R(w)}{r^2(1-w^{-4})}dr^2,
\end{equation}
\end{small}
where R is the Ad$S_5$ space radius and $r$ denotes the radial coordinate of the black brane
geometry. $x^i (i=1, 2, 3) $ is the boundary coordinate. Besides
\begin{small}
\begin{equation}
\begin{split}
T(w) &=1-k\Bigg(75w^{-4}+\frac{1225}{16}w^{-8}-\frac{695}{16}w^{-12}\Bigg)+...\\
X(w) &=1-\frac{25k}{16}w^{-8}(1+w^{-4})+...\\
R(w) &=1+k\Bigg(75w^{-4}+\frac{1175}{16}w^{-8}-\frac{4585}{16}w^{-12}\Bigg)+...
\end{split},
\end{equation}
\end{small}
and $w=\frac{r}{r_h}$ with $r=r_h$ being the event horizon and $r=\infty$ the boundary.
Further, $\lambda$ can be related to the ratio of the shear viscosity to the entropy density
$\eta \slash s$ by \cite{Buc1,Buc2,Mye}
\begin{small}
\begin{equation}
\frac{\eta}{s}=\frac{1}{4\pi}\Bigg(1+\frac{15\zeta(3)}{\lambda^{3/2}}\Bigg).
\end{equation}
\end{small}

The temperature of the black hole is
\begin{small}
\begin{equation}
T=\frac{r_h}{\pi R^2(1-k)}.
\end{equation}
\end{small}

\section{Potential analysis}
Following the calculations of \cite{Sat}  to study the Schwinger effect for
the background metric of (8), the Nambu-Goto string action is
\begin{small}
\begin{eqnarray}
S = T_F\int d\tau d\sigma \mathcal{L}=T_F\int d\tau d\sigma \sqrt{det G_{ab}},\qquad
G_{ab} \equiv \frac{\partial x^\mu}{\partial\sigma^a}\frac{\partial x^\nu}{\partial\sigma^b}g_{\mu\nu}
\end{eqnarray}
\end{small}
where $T_F=\frac{1}{2\pi\alpha^\prime}$ is the fundamental string tension,
and $G_{ab}(a,b=0,1)$ is the induced metric on the string world sheet,
using the static gauge
\begin{equation}
x^0=\tau,\quad x^1=\sigma,\quad x^2=x^3=const.
\end{equation}

Assuming the radial direction depends only on $\sigma$
\begin{equation}
r=r(\sigma).
\end{equation}

The metric(8) is calculated as
\begin{small}
\begin{equation}
ds^2=-\frac{r^2(1-w^{-4})}{R^2}T(w)dt^2+\Bigg(\frac{r^2X(w)}{R^2}+\frac{R^2}{r^2}\frac{R(w)}{(1-w^{-4})}\dot{r}^2\Bigg)d\sigma^2.
\end{equation}
\end{small}

The induced metric $G_{ab}$ is obtained
\begin{small}
\begin{equation}
G_{ab}=\left(
\begin{array}{ccc}
\frac{r^2(1-w^{-4})}{R^2}T(w)&0\\
0&\frac{r^2}{R^2}X(w)\!+\!\frac{R^2}{r^2}\frac{R(w)}{(1-w^{-4})}\dot{r}^2\\
\end{array}
\right),
\end{equation}
\end{small}
where $\dot{r}=\frac{\partial r}{\partial \sigma}$. Then $\mathcal{L}$ is
expressed as
\begin{small}
\begin{equation}
\mathcal{L} =\sqrt{detG_{ab}}
            =\sqrt{\frac{r^4}{R^4}(1-w^{-4})T(w)X(w)+\dot{r}^2T(w)R(w)}.
\end{equation}
\end{small}

Now $\mathcal{L}$ does not depend on $\sigma$ explicitly; then, the
corresponding Hamiltonian is a constant, which is

\begin{small}
\begin{equation}
\mathcal{H}=\mathcal{L}-\frac{\partial\mathcal{L}}{\partial\dot{r}}\dot{r}.
\end{equation}
\end{small}

Imposing the boundary condition at $\sigma=0$,
\begin{small}
\begin{equation}
\frac{dr}{d\sigma}=0 \quad r=r_c \quad  (r_h<r_c<r_0),
\end{equation}
\end{small}
the conserved quantity becomes

\begin{small}
\begin{equation}
\frac{\frac{r^4}{R^4}(1-w^{-4})T(w)X(w)}{\sqrt{\frac{r^4}{R^4}(1-w^{-4})T(w)X(w)+
\dot{r}^2T(w)R(w)}}=const
\equiv\sqrt{\frac{r_c^4}{R^4}(1-w_c^{-4})T(w_c)X(w_c)},
\end{equation}
\end{small}
with
\begin{small}
\begin{equation}
\begin{split}
T(w_c) &=1-k\Bigg(75\big(\frac{r_h}{r_c}\big)^4+
\frac{1225}{16}\big(\frac{r_h}{r_c}\big)^8-
\frac{695}{16}\big(\frac{r_h}{r_c}\big)^{12}\Bigg)+...\\
X(w_c) &=1-\frac{25k}{16}\big(\frac{r_h}{r_c}\big)^8
\Bigg(1+\big(\frac{r_h}{r_c}\big)^4\Bigg)+....
\end{split}
\end{equation}
\end{small}

From the conserved quantity, we obtained the $\dot{r}$
\begin{small}
\begin{equation}
\frac{dr}{d\sigma}=\frac{r^2}{R^2}
\sqrt{\frac{(1-w^{-4})X(w)}{R(w)}
\Bigg[\frac{r^4(1-w^{-4})T(w)X(w)}{r_c^4(1-w_c^{-4})T(w_c)X(w_c)}-1\Bigg]}.
\end{equation}
\end{small}

By integrating the expression (22), the separation distance x of test
particles on the probe brane is represented by
\begin{small}
\begin{equation}
x=\frac{2R^2}{ar_0}\int_1^{1/a}\frac{dy}{y^2\sqrt{1-w^{-4}}}
\sqrt{\frac{R(w)}{X(w)}}\frac{1}
{\sqrt{\frac{y^4(1-w^{-4})T(w)X(w)}{(1-w_c^{-4})T(w_c)X(w_c)}-1}},
\end{equation}
\end{small}
with
\begin{small}
\begin{equation}
\begin{split}
T(w_c) &=1-k\Bigg(75\big(\frac{b}{a}\big)^4+\frac{1225}{16}\big(\frac{b}{a}\big)^8-
\frac{695}{16}\big(\frac{b}{a}\big)^{12}\Bigg)+...\\
X(w_c) &=1-\frac{25k}{16}\big(\frac{b}{a}\big)^8\Bigg(1+\big(\frac{b}{a}\big)^4\Bigg)+...,
\end{split}
\end{equation}
\end{small}
where we have defined dimensionless quantities
\begin{small}
\begin{equation}
y\equiv\frac{r}{r_c},\quad a\equiv\frac{r_c}{r_0},\quad b\equiv\frac{r_h}{r_0}.
\end{equation}
\end{small}

The sum of the Coulomb potential(CP) and static energy(SE) in the $R^4$
correction is derived

\begin{small}
\begin{equation}
V_{CP+SE}=2T_F\int_0^{x/2}d\sigma\mathcal{L}
= 2T_Fr_0a\int_1^{1/a} dy\sqrt{T(w)R(w)}\times
    \frac{1}{\sqrt{1-\frac{(1-w_c^{-4})T(w_c)X(w_c)}{y^4(1-w^{-4})T(w)X(w)}}}.
\end{equation}
\end{small}

Next we calculate the critical electric field. The DBI action is given by
\begin{small}
\begin{equation}
\begin{split}
S_{DBI} &=-T_{D3}\int d^4x\sqrt{-det(G_{\mu\nu}+\mathcal{F}_{\mu\nu})} \\
\quad T_{D3} &=\frac{1}{g_s(2\pi)^3\alpha^{\prime2}}
\end{split},
\end{equation}
\end{small}
where $T_{D3}$ is the D3-brane tension.

As the virtue shows in Eq.(8), the induced metric $G_{\mu\nu}$ reads
\begin{small}
\begin{equation}
\begin{split}
G_{00} &=-\frac{r^2}{R^2}(1-w^{-4})T(w) \quad G_{11} =\frac{r^2}{R^2}X(w)\\
G_{22} &=\frac{r^2}{R^2}X(w)  \quad \quad \quad \quad \quad \quad G_{33} =\frac{r^2}{R^2}X(w)\\
\end{split}
\end{equation}
\end{small}

\begin{figure*}[htp]
\centering
\includegraphics[width=8.5cm]{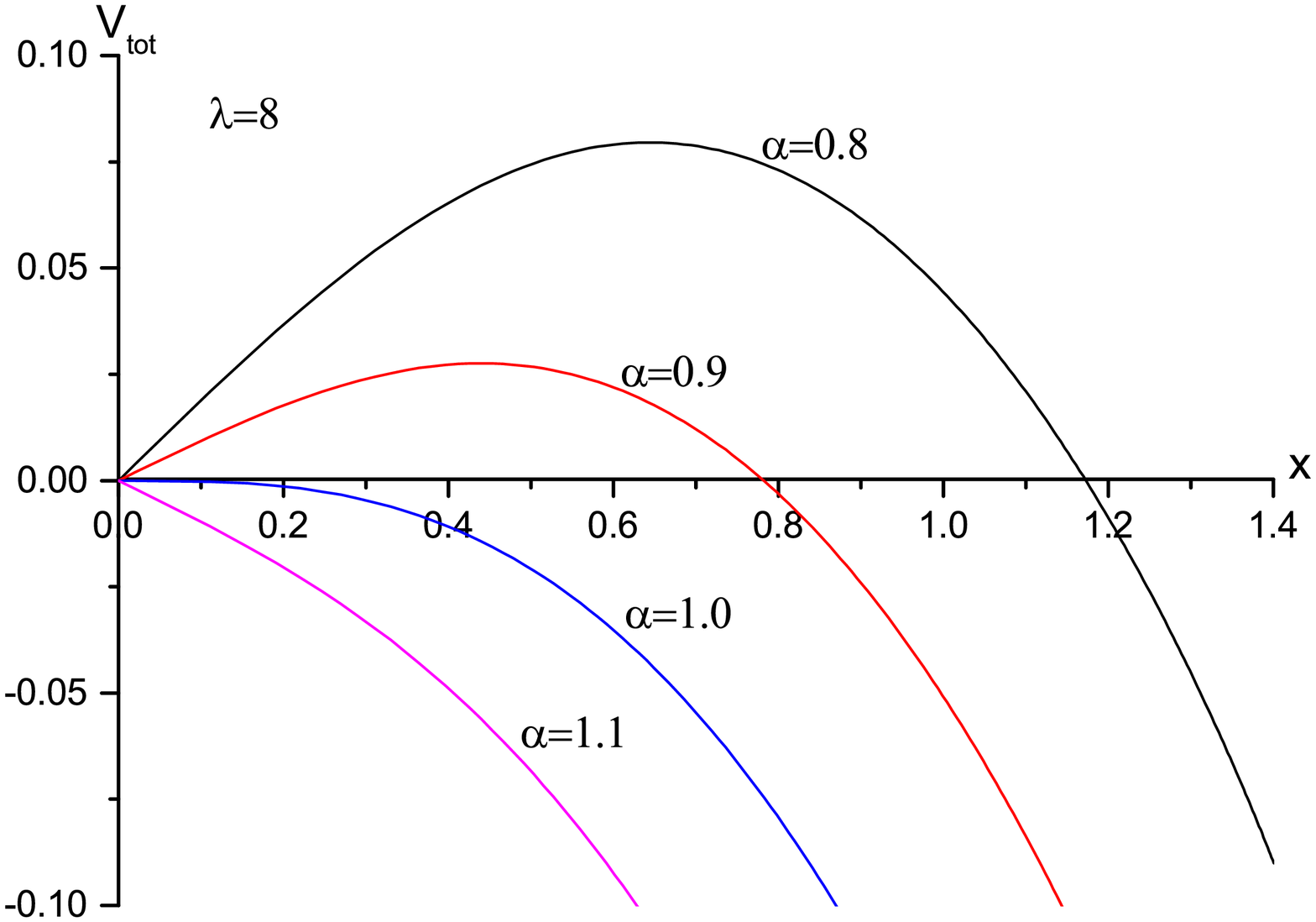}
\includegraphics[width=8.5cm]{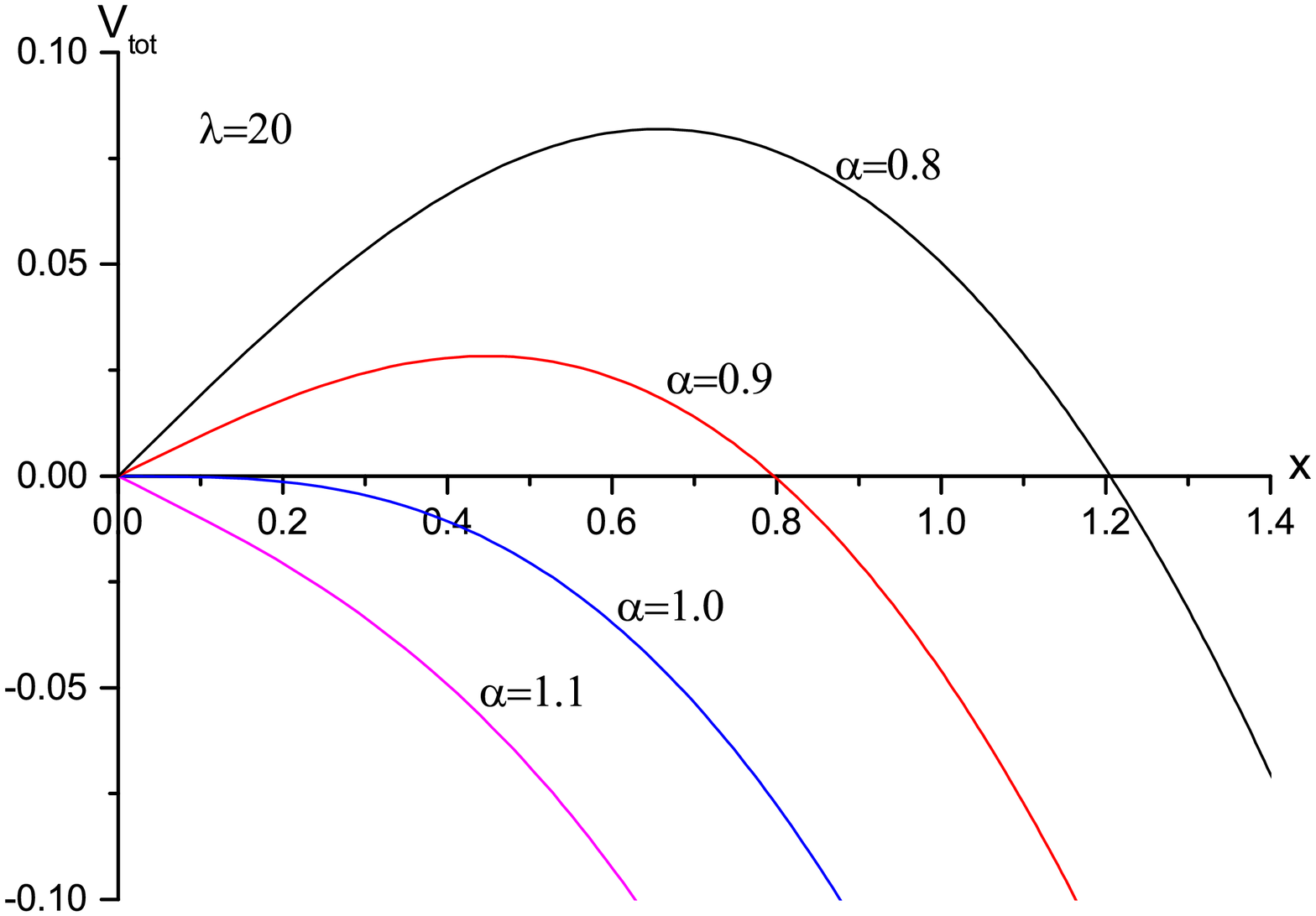}
\caption{$V_{tot}$ against x in the presence of $R^4$ corrections.
Left: $\lambda$ = 8; Right: $\lambda$ = 20. In all of the plots from top to bottom
$\alpha$ = 0.8, 0.9, 1.0, 1.1, respectively. }
\label{label}
\end{figure*}

According to \cite{Zwi}, $\mathcal{F}_{\mu\nu}=2\pi\alpha^\prime F_{\mu\nu}$,
one obtains
\begin{small}
\begin{equation}
\begin{split}
&G_{\mu\nu}+\mathcal{F}_{\mu\nu}\\
&=\left(
\begin{array}{cccc}
-\frac{r^2(1-w^{-4})}{R^2}T(w)&2\pi\alpha^\prime E&0&0\\
-2\pi\alpha^{\prime}E&\frac{r^2}{R^2}X(w)&0&0\\
0&0&\frac{r^2}{R^2}X(w)&0\\
0&0&0&\frac{r^2}{R^2}X(w)\\
\end{array}
\right),
\end{split}
\end{equation}
\end{small}
which yields
\begin{small}
\begin{equation}
det(G_{\mu\nu}+\mathcal{F}_{\mu\nu})
=-\big(\frac{r}{R}\big)^4X^2(w)
\Bigg[\frac{r^4(1-w^{-4})}{R^4}T(w)X(w)-(2\pi\alpha^\prime E)^2\Bigg],
\end{equation}
\end{small}
where we consider the total static-potential by turning on an electric field
E along the $x^1$-direction \cite{Sat2}.

Combining (30) with (27) and setting the D3-brane at $r=r_0$, one gets
\begin{small}
\begin{equation}
S_{DBI}= -T_{D3}\frac{r_0^4}{R^4}
 \cdot\int d^4x\sqrt{X^2(w_0)\Bigg[(1-w_0^{-4})T(w_0)X(w_0)-
\frac{(2\pi\alpha^\prime)^2 E^2 R^4}{r_0^4}\Bigg]},
\end{equation}
\end{small}
with

\begin{small}
\begin{equation}
\begin{split}
T(w_0) &=1-k\Bigg(75b^4+\frac{1225}{16}b^8-\frac{695}{16}b^{12}\Bigg)+...,\\
X(w_0) &=1-\frac{25k}{16}b^8(1+b^4)+...,
\end{split}
\end{equation}
\end{small}
where $w_0=r_0/r_c=1/b$.

It is clear that the square root in (31) is non-negative
\begin{small}
\begin{equation}
(1-w_0^{-4})T(w_0)X(w_0)-\frac{(2\pi\alpha^\prime)^2 E^2 R^4}{r_0^4}\geq0,
\end{equation}
\end{small}
which leads to
\begin{small}
\begin{equation}
E_c\leq T_F\frac{r_0^2}{R^2}\sqrt{(1-w_0^{-4})T(w_0)X(w_0)}.
\end{equation}
\end{small}

It is clear that the critical field $E_c$ in the AdS black hole background
with $R^4$ correction equals
\begin{small}
\begin{equation}
E_c=T_F\frac{r_0^2}{R^2}\sqrt{(1-b^4)T(w_0)X(w_0)}.
\end{equation}
\end{small}

Next, we define a dimensionless electric field $\alpha$ that measures E in
a unit of the critical field $E_c$,
\begin{small}
\begin{equation}
\alpha\equiv\frac{E}{E_c}.
\end{equation}
\end{small}

With the electrostatic potential E,one finds the total potential $V_{tot}$ as
\begin{small}
\begin{equation}
\begin{split}
V_{tot}&=V_{CP+SE}-Ex\\
       &=2T_Fr_0\Bigg[a\int_1^{1/a}dy\frac{\sqrt{T(w)R(w)}}
       {\sqrt{1-\frac{(1-w_c^{-4})T(w_c)X(w_c)}{y^4(1-w^{-4})T(w)X(w)}}}\\
       &-\frac{\alpha}{a}\sqrt{(1-b^4)T(w_0)X(w_0)}
       \times\int_1^{1/a}\frac{dy}{y^2\sqrt{1-w^{-4}}}\sqrt{\frac{R(w)}{X(w)}}\frac{1}
       {\sqrt{\frac{y^4(1-w^{-4})T(w)X(w)}{(1-w_c^{-4})T(w_c)X(w_c)}-1}}\Bigg].
\end{split}
\end{equation}
\end{small}

For comparing with the Einstein case in \cite{Sat}, we set the fixed temperature
$b=0.5$. In Fig.1, we plot the total potential $V_{tot}$ as a function of the
inter-distance $x$ with coupling parameter $\lambda=8$ (left panel) and $\lambda=20$
(right panel). The shapes of the potential are plotted for $\alpha=0.8,0.9,1.0$ and
1.1 from top to bottom, where $T_Fr_0=L^2/r_0=1$. From the figures, we can see that
the potential barrier vanishes for $\alpha\geq 1.0$ and the critical field is
$\alpha=1.0$($E=E_c$), which is in agreement with the DBI result.

\begin{figure*}[htp]
\centering
\includegraphics[width=8.5cm]{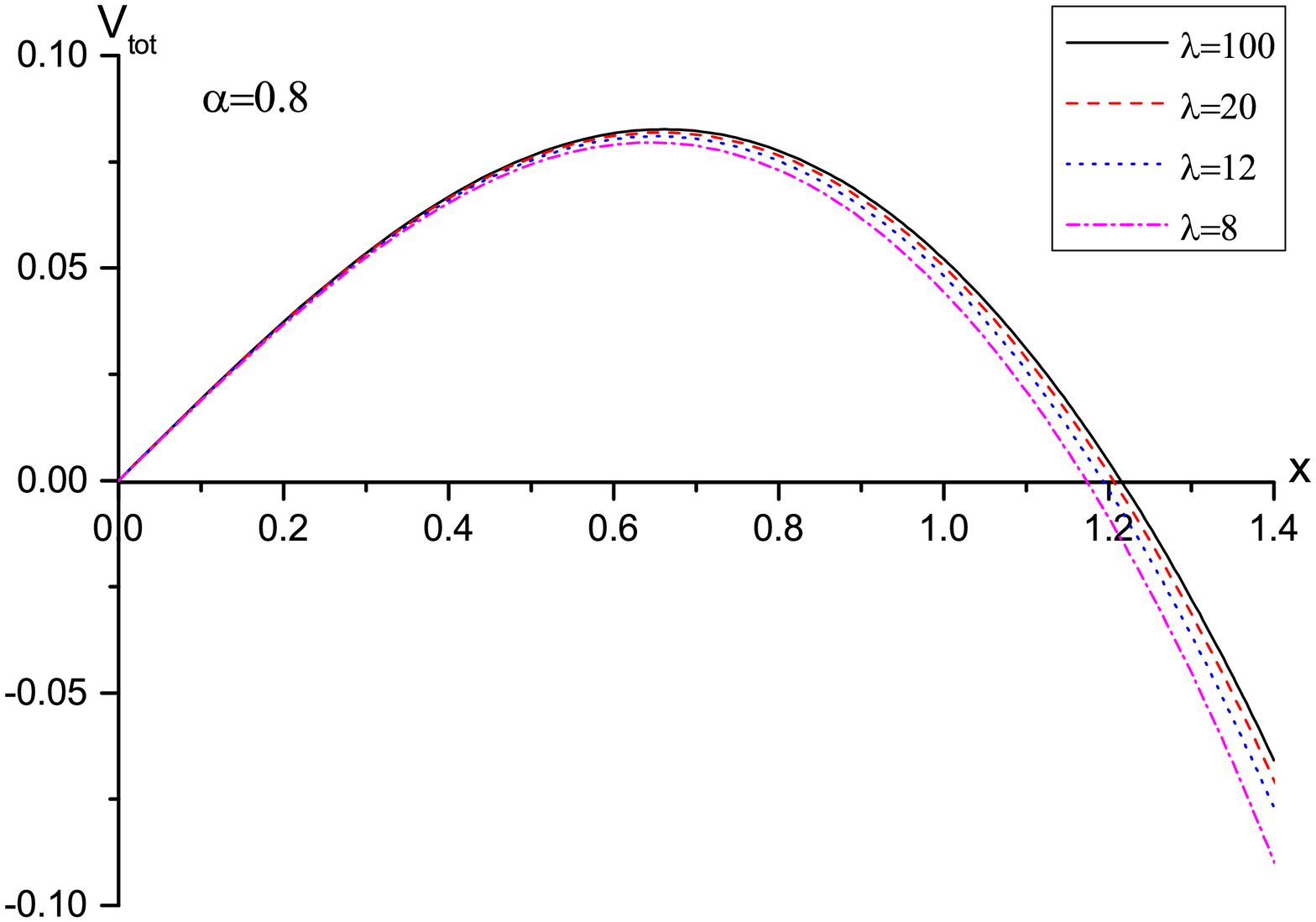}
\includegraphics[width=8.5cm]{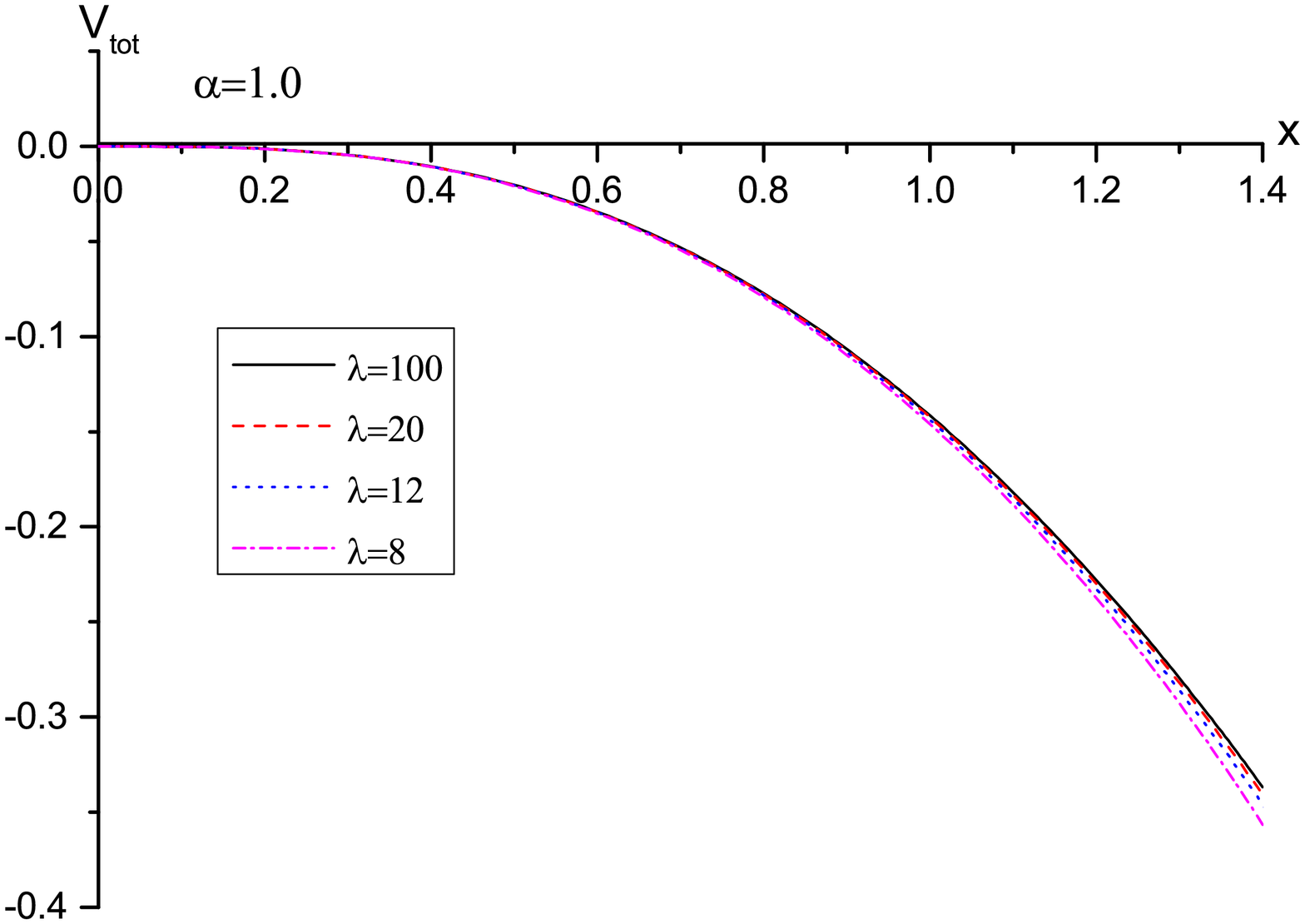}
\caption{$V_{tot}$ against x in the presence of $R^4$ corrections.
Left: $\alpha = 0.8$; Right: $\alpha = 1.0$.
In all of the plots from top to bottom $\lambda = 100, 20, 12, 8$, respectively. }
\label{label}
\end{figure*}

To study the effect of $R^4$ corrections on the potential barrier, we plot $V_{tot}$
against x at $\alpha =0.8$, with different values of $\lambda$ in the left panel of Fig.2.
It is found that decreasing $\lambda$ leads to decreasing height and width of the barrier,
and it is known that a higher barrier means the more difficult the pair production.
This proves a way to enhance the Schwinger effect by decreasing the value of $\lambda$.

Further, we plot the shape of $V_{tot}$ against x at $\alpha = 1.0$ with different $\lambda$
in the right panel of Fig.2, to show the effect of $R^4$ corrections on $E_{c}$. It is implies
that the vacuum becomes unstable when the barrier starts to vanish (at $\alpha \geq 1.0$) for each $\lambda$.

These results are consistent with the anticipation of the expression of production rate in section-$\mathbf{1}$.

\section{Confining D3-brane background}
In this section, we study the Schwinger effect in a confining D3-brane background with $R^4$
corrections. The induced metric is \cite{Kaw1}
\begin{small}
\begin{equation}
ds^2=-\frac{r^2 T(w)dt^2}{R^2}+\frac{r^2 X(w)}{R^2}\Bigg[(dx^1)^2+(dx^2)^2
+(1-w^{-4})(dx^3)^2\Bigg]+\frac{R^2 R(w)dr^2}{(1-w^{-4})r^2},
\end{equation}
\end{small}
with
\begin{small}
\begin{equation}
\begin{split}
T(w) &=1-k\Bigg(75w^{-4}+\frac{1225}{16}w^{-8}-\frac{695}{16}w^{-12}\Bigg)+...\\
X(w) &=1-\frac{25k}{16}w^{-8}(1+w^{-4})+...\\
R(w) &=1+k\Bigg(75w^{-4}+\frac{1175}{16}w^{-8}-\frac{4585}{16}w^{-12}\Bigg)+...
\end{split},
\end{equation}
\end{small}
and $r_h$ represents the inverse compactification radius in the $x^3$-direction.

Taking the previous steps, we obtained the inter-distance $\emph{x}$ and the sum of potential
energy and static energy $V_{CP+SE}$, respectively. One goes
\begin{small}
\begin{equation}
x=\frac{2R^2}{ar_0}\int_1^{1/a}\sqrt{\frac{R(w)}{X(w)(1-w^{-4})}}\frac{dy}
{y^2\sqrt{\frac{y^4T(w)X(w)}{T(w_c)X(w_c)}-1}},
\end{equation}
\end{small}

\begin{small}
\begin{equation}
V_{CP+SE}=2T_Far_0
\int_1^{1/a}\sqrt{\frac{T(w)R(w)}{(1-w^{-4})}}\frac{y^2dy}
{\sqrt{y^4-\frac{T(w_c)X(w_c)}{T(w)X(w)}}}.
\end{equation}
\end{small}

\begin{figure*}[htp]
\centering
\includegraphics[width=8.5cm]{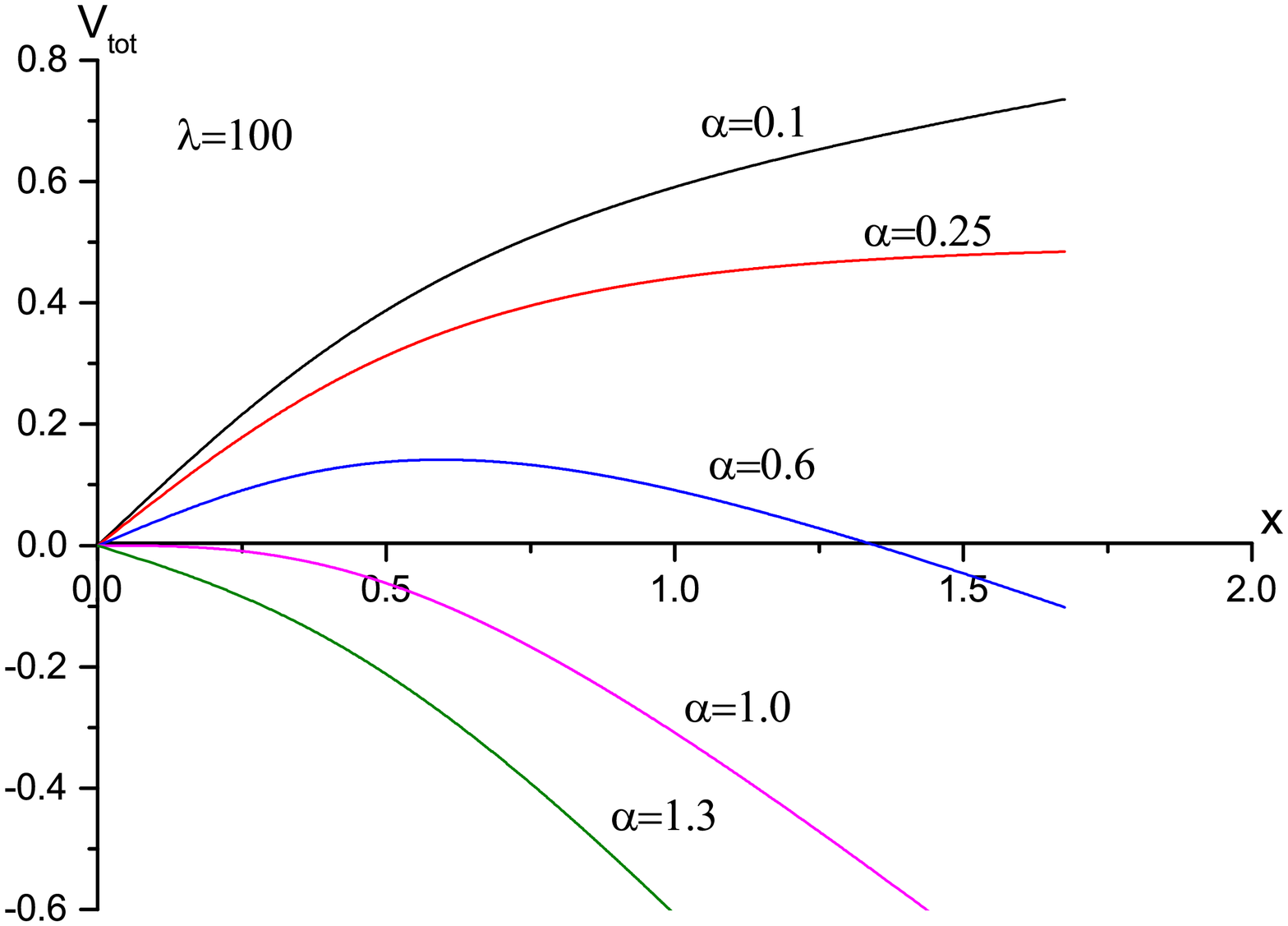}
\includegraphics[width=8.5cm]{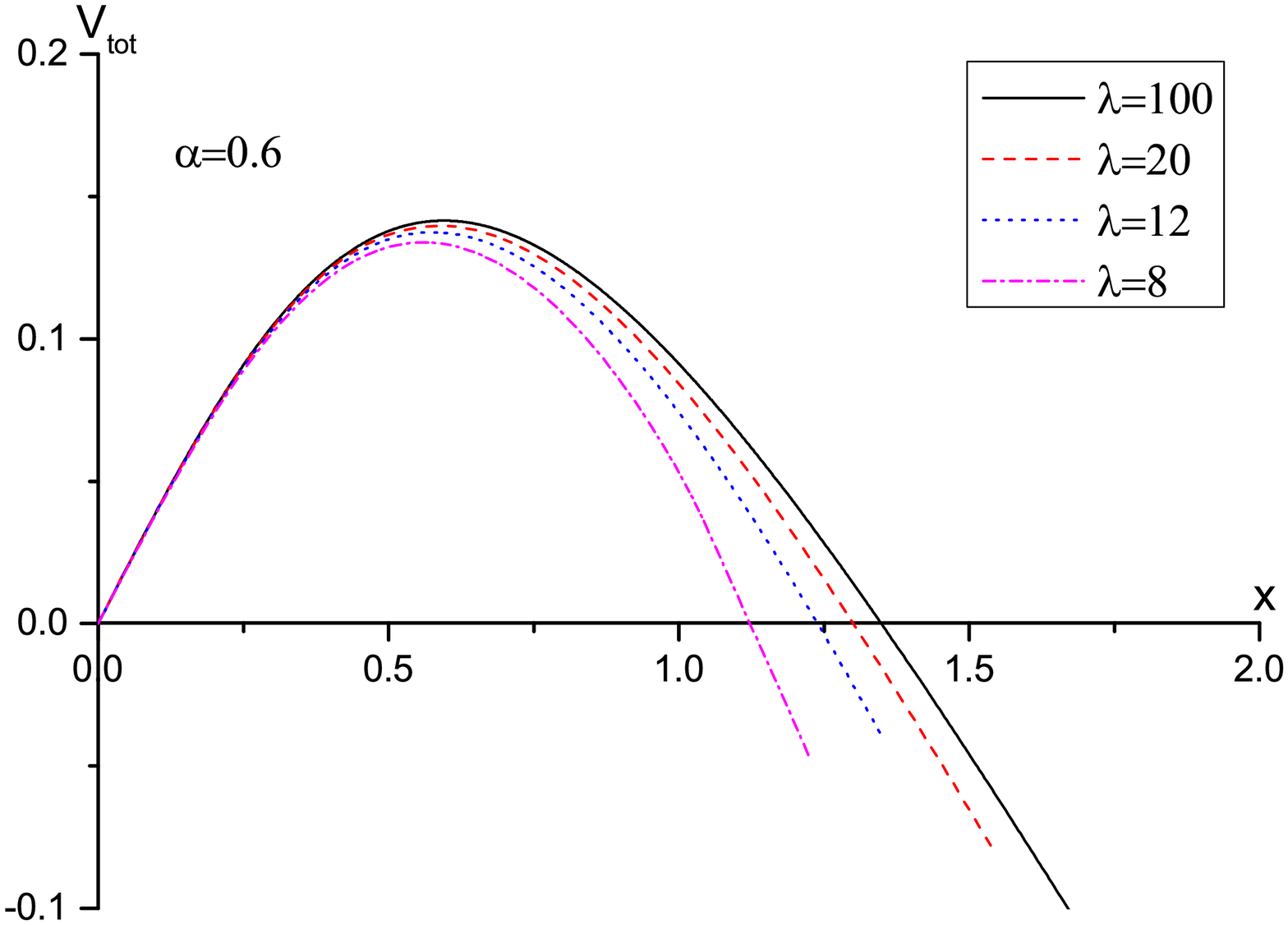}
\caption{$V_{tot}$ against $X$. Left: $\lambda = 100$, from top to bottom
$\alpha=0.1, 0.25, 0.6, 1.0, 1.3$; Right: $\alpha=0.6$, from top to bottom
$\lambda=100,20,12,8$, respectively. }
\label{label}
\end{figure*}

where $T(w_c) $ and $X(w_c)$ is defined in (24).

Now, we study the critical electric field as we did before, and the induced metric is :

\begin{equation}
\begin{split}
G_{00}&=-\frac{r^2}{R^2}T(w) \quad \quad \ G_{11}=\frac{r^2}{R^2}X(w)\\
G_{22}&= \frac{r^2}{R^2}X(w) \qquad \quad G_{33}=\frac{r^2}{R^2}(1-w^{-4})X(w)
\end{split}
\end{equation}

Then we obtain
\begin{small}
\begin{equation}
\begin{split}
G_{\mu\nu}+\mathcal{F}_{\mu\nu}=
&\left(
\begin{array}{cccc}
-\frac{r^2}{R^2}T(w)&2\pi\alpha^\prime E&0&0\\
-2\pi\alpha^{\prime}E&\frac{r^2}{R^2}X(w)&0&0\\
0&0&\frac{r^2}{R^2}X(w)&0\\
0&0&0&\frac{r^2(1-w^{-4})}{R^2}X(w)\\
\end{array}
\right)
\end{split}
\end{equation}
\end{small}
which yields
\begin{small}
\begin{equation}
det(G_{\mu\nu}+\mathcal{F}_{\mu\nu})
=-\frac{r^4}{R^4}X^2(w)(1-w^{-4})
\Bigg[\frac{r^4}{R^4}T(w)X(w)-(2\pi\alpha^\prime E)^2\Bigg],
\end{equation}
\end{small}
where we assume the electric field E is turned on along the $x^1$ -direction as well.

Substituting (44) into (27) and setting the D3-brane at $r=r_0$, one can obtain
\begin{small}
\begin{equation}
S_{DBI}=-T_{D3}\frac{r_0^4}{R^4}
\int d^4x\sqrt{X^2(w_0)(1-w_0^{-4})
\Bigg[T(w_0)X(w_0)-\frac{(2\pi\alpha^\prime)^2 E^2 R^4}{r_0^4}\Bigg]},
\end{equation}
\end{small}
then, obviously,
\begin{small}
\begin{equation}
X^2(w_0)(1-w_0^{-4})>0.
\end{equation}
\end{small}

For ensuring the square root in (45) is non-negative
\begin{small}
\begin{equation}
T(w_0)X(w_0)-\frac{(2\pi\alpha^\prime)^2 E^2 R^4}{r_0^4}\geq0,
\end{equation}
\end{small}
which means
\begin{small}
\begin{equation}
E\leq T_F\frac{r_0^2}{R^2}\sqrt{T(w_0)X(w_0)}.
\end{equation}
\end{small}

\begin{figure*}[htp]
\centering
\includegraphics[width=8.5cm]{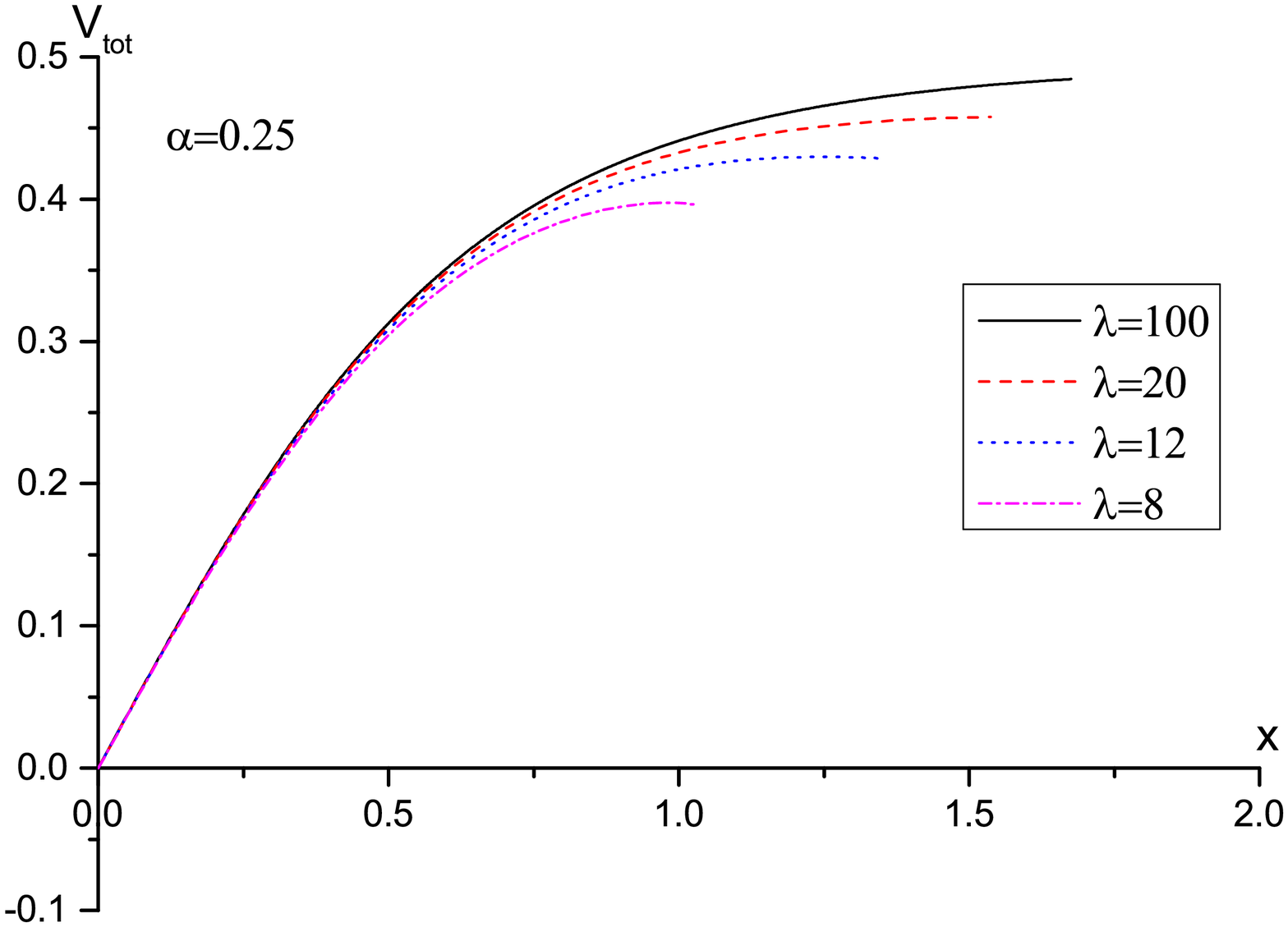}
\includegraphics[width=8.5cm]{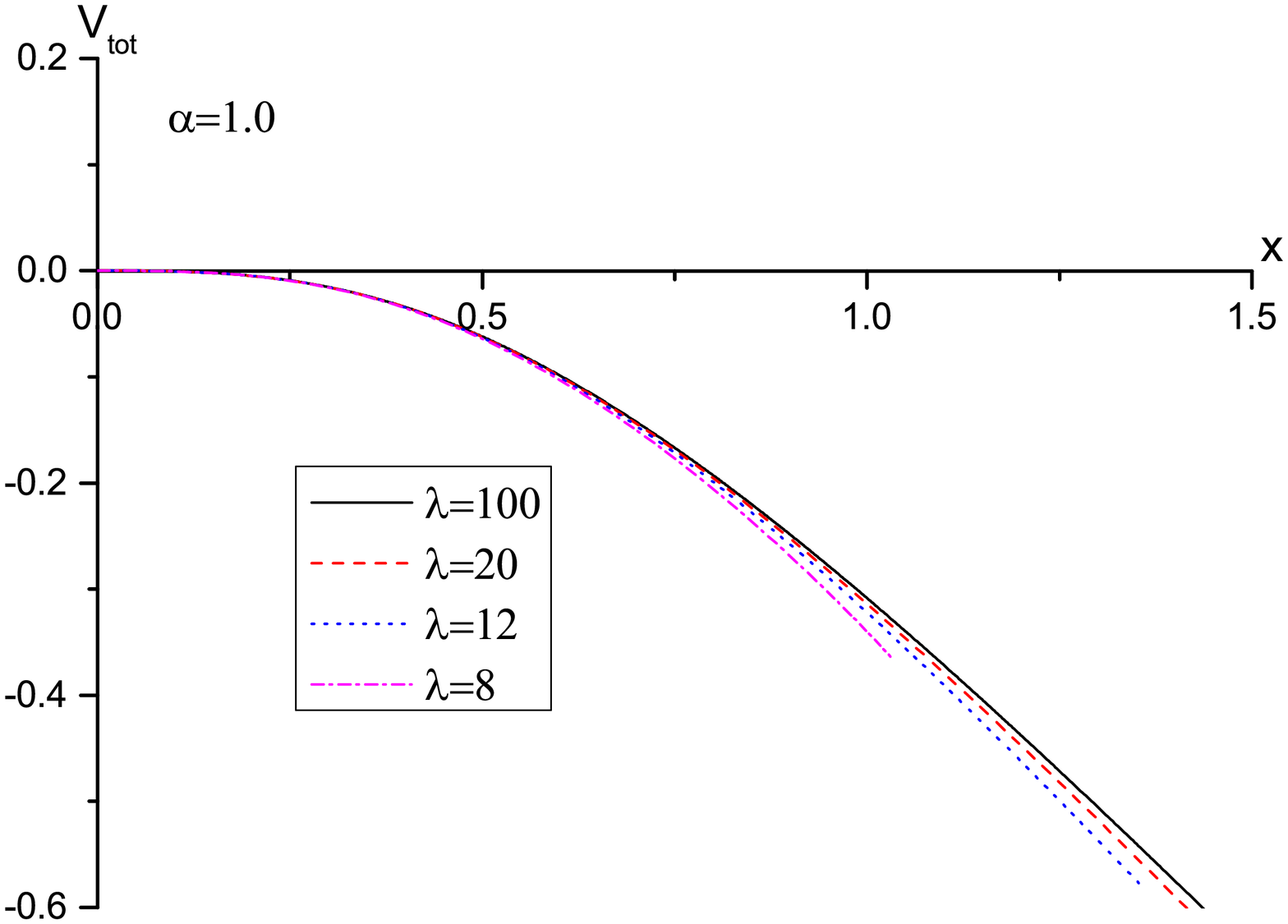}
\caption{$V_{tot}$ against $X$. Left: $\alpha=0.25$; Right: $\alpha=1.0$.
In all of the plots from top to bottom $\lambda=100, 20, 12, 8$, respectively. }
\label{label}
\end{figure*}
Obviously, the critical field $E_c$ in the confining D3-brane background with $R^4$
corrections is obtained
\begin{small}
\begin{equation}
E=T_F\frac{r_0^2}{R^2}\sqrt{T(w_0)X(w_0)}.
\end{equation}
\end{small}

The total potential is
\begin{small}
\begin{equation}
\begin{split}
V_{tot}&=V_{CP+SE}-Ex\\
       &=2T_Fr_0\Bigg[a\int_1^{1/a}\sqrt{\frac{T(w)R(w)}{(1-w^{-4})}}
       \frac{y^2dy}{\sqrt{y^4-\frac{T(w_c)X(w_c)}{T(w)X(w)}}}\\
       &-\quad\frac{\alpha}{a}\sqrt{T(w_0)X(w_0)}
       \times\int_1^{1/a}
       \sqrt{\frac{R(w)}{X(w)(1-w^{-4})}}\frac{dy}{y^2\sqrt{\frac{y^4T(w)X(w)}{T(w_c)X(w_c)}-1}}\Bigg].
\end{split}
\end{equation}
\end{small}
where $\alpha$ is defined in (36).

We plot $V_{tot}$ versus x with $\lambda =100$ by setting b = 0.5 and $2L^2/{r_0} = 2T_Fr_0=1$
in the left panel of Fig.3, the same as \cite{Sat2}, which is similar picture to other cases with different
$\lambda$. As the figures shows, there exist two critical values for the electric field: one is at
$\alpha =1$($E = E_c$), the other is at $\alpha = 0.25$($E = E_s = 0.25E_c$).

As we did before, to analyze the effect of $R^4$ corrections on the potential barrier, we plot
$V_{tot}$ versus x at $\alpha = 0.6$ with different $\lambda$ in the right panel of Fig.3,
which shows that as $\lambda$ decreases both the height and width of the potential barrier
decrease. Therefore, one could conclude that increasing $\lambda$ enhances the Schwinger effect.

Similarly, to show the effect of $R^4$ corrections on the two critical fields, we plot $V_{tot}$
versus x at $\alpha = 0.25$ and $\alpha = 1$ with different $\lambda$ in Fig.4. From the left
panel, one can see that for various $\lambda$, the potential with $\alpha = 0.25$ will saturate
at large x. From the right panel, one can find that the barrier vanishes for each plot at $\alpha =1$,which is in agreement with the DBI result.

These results are consistent with our theoretical expectations in section-$\mathbf{1}$.

\section{Conclusion and discussion}
We have investigated $R^4$ corrections to the holographic Schwinger effect in an AdS black hole
background and a confining D3-brane background. The critical value for the electric
field is obtained. It is shown that for both backgrounds, decreasing the parameter $\lambda$
enhances the Schwinger effect.

Further, the study on the relation between Schwinger effect and $\eta/s$ at strong coupling, as
expressed in Eq.(10), shows that increasing $\lambda$ leads to decreasing $\eta/s$
thus making the fluid conform more to the ideal case, and decreasing $\lambda$ leads to increasing
Schwinger effect. Thus, one can conclude that the Schwinger effect is enhanced while the $\eta/s$
decreases at strong coupling.

\acknowledgments{Finally, we acknowledge the financial support from NSFC (11705166 ,11475149).
The authors thank PH.D. Yi-long Xie and Di-kai Li for improving the English.}


\end{document}